\documentclass[12pt]{article}
\usepackage[numbers]{natbib}
\usepackage{a4}
\usepackage{amsmath}
\usepackage{amssymb}
\usepackage{latexsym}
\usepackage{amssymb}
\usepackage{epsfig}
\usepackage{natbib}
\def \TT{{\mathrm{T}}}

\def \d{{\mathrm{d}}}

\def \pd{\partial}

\def \tl#1{\overset{\kern 1pt\circ}{#1}}
\def \TL#1{\overset{\kern -3pt \circ}{#1}}
\def \TLL#1{\overset{\kern -7pt \circ}{#1}}

\def \Bbeta{\boldsymbol{\beta}}

\def \Bx{{\boldsymbol{x}}}
\def \Bv{{\boldsymbol{v}}}

\def \LL{{\cal{L}}}

\begin{document}
\title{{\bf The gauge theory of dislocations: 
a nonuniformly moving screw dislocation}}
\author{
Markus Lazar~$^\text{a,b,}$\footnote{
{\it E-mail address:} lazar@fkp.tu-darmstadt.de (M.~Lazar).} 
\\ \\
${}^\text{a}$ 
        Emmy Noether Research Group,\\
        Department of Physics,\\
        Darmstadt University of Technology,\\
        Hochschulstr. 6,\\      
        D-64289 Darmstadt, Germany\\
${}^\text{b}$ 
Department of Physics,\\
Michigan Technological University,\\
Houghton, MI 49931, USA
}

\date{\today}    
\maketitle

\begin{abstract}
We investigate the nonuniform motion of a straight screw dislocation
in infinite media in the framework of the translational
gauge theory of dislocations.
The equations of motion are derived for an arbitrarily moving screw
dislocation. The fields of the 
elastic velocity, elastic distortion, dislocation density and dislocation
current surrounding the arbitrarily moving screw dislocation
are derived explicitely in the form of integral representations. 
We calculate the radiation fields and the fields depending on the 
dislocation velocities.
\\

\end{abstract}

\section{Introduction}

For many years,
the investigation of the nonuniform motion of dislocations 
has attracted the interest of researchers
in different fields such as physics, material science, continuum mechanics, seismology 
and earthquake engineering. 
Usually, the motion of dislocations is investigated in the framework of 
incompatible
elastodynamics where the dislocation density tensor 
and the dislocation current tensor 
are given as source terms of the elastic fields 
(see, e.g., \citet{Mura63,Mura,Lardner,Gunther73}).
The behaviour of the motion of a 
dislocation is somehow particular, because at any time
the fields are determined not only by the instantaneous values of 
the velocity (or higher derivatives of the position with respect to time),
but also by the values of these quantities in the past~\citep{Eshelby51,Eshelby53}. 
As \citet{Eshelby51} succinctly put it: `The dislocation is haunted by its past'.
This fact is based on the property that Huygens' principle is not
valid in two dimensions~(see, e.g., ~\citep{Wl}).
Solutions for nonuniformly moving screw dislocations have been given by
\citet{Nabarro51,Eshelby51,Eshelby53,KM64}, and \citet{XM80}. Results for the nonuniform
motion of a gliding edge dislocation have been given 
by~\citet{Mura64,XM81}, and \citet{Brock82}. 

Immediately, the question comes up, what is the suitable theory for an improved 
approach of a continuum theory of dislocation dynamics.
Generalized theories of elasticity are, for instance, 
nonlocal elasticity~\citep{Eringen83,Eringen02}, 
gradient elasticity~\citep{Mindlin64,Mindlin72} 
and dislocation gauge theory~\citep{Edelen83,Edelen88,Lazar00,LA08}.
However, in the dynamical version of nonlocal elasticity it is not obvious 
which dynamical kernel should be used. In the dynamical extension of gradient 
elasticity~\citep{Aifantis06,Aifantis07} 
the choice of `dynamical' gradients and length scales 
is based on ad-hoc assumptions.
On the other hand, dislocation gauge theory is not based on such assumptions.
The choice of the dynamical state variables is given by a canonical field theoretical framework~\citep{LA08}.
For this reason, dislocation gauge theory seems to be the appropriate theory
and it will be used in this paper.

Thus, a very promising and straightforward 
candidate for an improved dynamical approach of dislocations
including scale effects 
is the translational gauge theory of
dislocations~\citep{Edelen83,Edelen88,LA08,LA09}. 
In the gauge theory of dislocations, 
dislocations arise naturally as a consequence
of broken translational symmetry and their existence is not required to be
postulated a priori.
Moreover, such a theory uses the field theoretical 
framework which is well accepted in 
theoretical physics. 
From the point of view of theoretical physics, the gauge field approach of 
dislocations~\citep{Edelen83,Edelen88,LA08,LA09}
removes the singularity at the core of
a moving dislocation at subsonic as well as at supersonic speed~\citep{Lazar09},
the gauge theoretical approach in essence being reminiscent of nonlocality. 
Up to now, no solution for a nonuniformly moving dislocation has been given
in the dislocation gauge theory or another generalized theory of elasticity. 
The aim of this paper is the examination of a nonuniformly moving 
screw dislocation, for the first time,
in the framework of dislocation gauge theory. 
We will calculate the retarded expressions of the elastic fields as well as
of the dislocation density and dislocation current tensors.
This paper provides new insights on the change of 
the dislocation core structure while the dislocation is moving at sound wave speed. 

\section{Gauge theory of dislocations}

In this section, we briefly review the gauge theory of dislocations in the form 
given by~\citet{LA08}, and \citet{Lazar09}.

In dislocation dynamics, the following state quantities of 
dislocations are of importance\footnote{We use the usual notations: $\beta_{ij,k}:=\pd_k \beta_{ij}$ and 
$\dot{\beta}_{ij}:=\pd_t \beta_{ij}$.} 
\begin{align}
 \label{disl-den}
T_{ijk}=\beta_{ik,j} - \beta_{ij,k},\qquad    
I_{ij}=-v_{i,j} + {\dot{\beta}}_{ij},
\end{align}  
which are called the dislocation density tensor and 
the dislocation current tensor, respectively. 
They are kinematical quantities of dislocations 
and they are given in terms of the incompatible 
elastic distortion tensor $\beta_{ij}$ and incompatible 
physical velocity of the material continuum $v_i$.
Their dimensions are:
$[\beta_{ij}]=1$, 
$[v_{i}]=\text{length}/\text{time}$,
$[T_{ijk}]=\text{1}/\text{length}$ and
$[I_{ij}]=\text{1}/\text{time}$.
The dislocation density and the dislocation current tensors fulfill 
the translational Bianchi identities
\begin{align}
\label{BI}
\epsilon_{jkl}T_{ijk,l}=0,\qquad 
\dot{T}_{ijk} + I_{ij,k}-I_{ik,j}= 0.    
\end{align}

In the dynamical translation gauge theory of dislocations, the Lagrangian
density is of the bilinear form 
\begin{align}
\label{tot-Lag}
\LL =T-W=
\frac{1}{2}\,p_{i}v_{i} 
+ \frac{1}{2}\,D_{ij}I_{ij}-\frac{1}{2}\,\sigma_{ij}\beta_{ij}  
- \frac{1}{4}\,H_{ijk}T_{ijk}.
\end{align}
The canonical conjugate quantities (response quantities) are defined by
\begin{align}
\label{can-qua}
p_i:=\frac{\pd \LL}{\pd v_i},\qquad
\sigma_{ij}:=-\frac{\pd \LL}{\pd \beta_{ij}} ,\qquad
D_{ij} :=\frac{\pd \LL}{\pd I_{ij}},\qquad
H_{ijk}:=-2\frac{\pd \LL}{\pd T_{ijk}},
\end{align}
where $p_i$, $\sigma_{ij}$, $D_{ij}$, and $H_{ijk}$ are the linear momentum vector,
the force stress tensor, 
the dislocation momentum flux tensor, and the pseudomoment stress
tensor\footnote{The moment stress tensor $\tau_{ijk}=-\tau_{jik}$ 
can be obtained from the 
pseudomoment stress tensor: $\tau_{ijk}=-H_{[ij]k}$ (see \citep{LA09,LH09}).},
respectively. They have the dimensions:
$[p_i]=\text{momentum}/(\text{length})^3
\stackrel{\rm SI}{=}\text{N\,s}/\text{m}^3$, 
$[D_{ij}]=\text{momentum}/(\text{length})^2
\stackrel{\rm SI}{=}\text{N\,s}/\text{m}^2$, 
$[\sigma_{ij}]=\text{force}/(\text{length})^2
\stackrel{\rm SI}{=}\text{Pa}$ and 
$[H_{ijk}]=\text{force}/\text{length}\stackrel{\rm SI}{=}\text{N}/\text{m}$. 

The Euler-Lagrange equations derived from the total Lagrangian density
$\LL=\LL(v_i,\beta_{ij},I_{ij},T_{ijk})$ are given by
\begin{align}
\label{euler-lag-2} 
&E^{\, \Bv}_i(\LL)= \pd_t  \frac{\pd \LL}{\pd \dot{v}_i} 
+ \pd_j\frac{\pd \LL}{\pd v_{i,j}} - \frac{\pd \LL}{\pd v_{i}} = 0,\\
\label{euler-lag-3}
&E^{\, \Bbeta}_{ij}(\LL)=\pd_t \frac{\pd \LL}{\pd \dot{\beta}_{ij}} 
+ \pd_k \frac{\pd \LL}{\pd \beta_{ij,k}} - \frac{\pd \LL}{\pd \beta_{ij}}  = 0.
\end{align}
We add to $\LL$ a so-called null Lagrangian, $\LL_N=\sigma^0_{ij}\beta_{ij}-p^0_i v_i$, 
with the `background' stress $\sigma^0_{ij}$
and the `background' momentum $p^0_i$ as external source fields, which satisfy:
$\dot{p}^0_i -  \sigma^0_{ij,j}= 0$.
In terms of the canonical conjugate quantities~(\ref{can-qua}),
Eqs.~(\ref{euler-lag-2}) and (\ref{euler-lag-3}) take the form
\begin{alignat}{2}
\label{inhom-di-1}
 D_{ij,j}+ p_i&=p^0_i,
\qquad &&(\text{momentum balance of dislocations}),\\
\label{inhom-di-2}
\dot{D}_{ij} + H_{ijk,k}+ \sigma_{ij}&=\sigma^0_{ij},
\qquad &&(\text{stress balance of dislocations}).
\end{alignat}
Eqs.~(\ref{inhom-di-1}) and (\ref{inhom-di-2})
represent the dynamical equations for the balance of dislocations.
Eq.~(\ref{inhom-di-1}) represents 
the momentum balance law of dislocations, where 
the physical momentum is the source of the dislocation momentum flux.
Eq.~(\ref{inhom-di-2}) represents the stress balance of dislocations.
The force stress and the time derivative of the dislocation momentum flux
are the sources of the pseudomoment stress.
It can be seen in Eqs.~(\ref{inhom-di-1}) and (\ref{inhom-di-2}) that the sources 
of the dislocation momentum flux and the pseudomoment stress tensors are the
so-called effective momentum vector $(p_i-p^0_i)$ 
and the effective force stress tensor
$(\sigma_{ij}-\sigma^0_{ij})$ (see, e.g., \cite{Edelen88}). 
These effective fields, which are the difference between the physical field and the 
background field, drive the dislocation fields.

The conservation of linear momentum  appears as an integrability condition 
from~(\ref{inhom-di-1}) and (\ref{inhom-di-2}).
It reads
\begin{alignat}{2}
\label{inhom-di}
\dot{p}_i -  \sigma_{ij,j}&= 0,
\qquad &&(\text{force balance of elasticity}),
\end{alignat}
where the time-derivative of the physical momentum vector 
is the source of the force stress tensor.

The linear, isotropic constitutive relations for the momentum, the force stress, 
the dislocation momentum flux and 
the pseudomoment stress are
\begin{align}
\label{con-eq}
p_i&=\rho v_i,\\
\label{con-eq-2}
\sigma_{ij}&= \lambda \delta_{ij} \beta_{kk} + \mu (\beta_{ij}+\beta_{ji}) + \gamma (\beta_{ij}-\beta_{ji}),\\
\label{con-eq-3}
D_{ij}&= d_1 \delta_{ij} I_{kk} + d_2 (I_{ij} + I_{ji}) + d_3 (I_{ij} - I_{ji}),\\
\label{con-eq-4}
H_{ijk}&= c_1 T_{ijk} + c_2 (T_{jki} - T_{kji}) + c_3 (\delta_{ij}T_{llk} - \delta_{ik}T_{llj}),
\end{align}
where $\rho$ is the mass density. 
Here $\mu, \lambda, \gamma$ are the elastic stiffness parameters,
$c_1,c_2,c_3$ denote the resistivity parameters associated with dislocations
and $d_1,d_2,d_3$ are the inertia terms associated with dislocation currents.
The nine material parameters have the dimensions:
$[\mu, \lambda, \gamma]=\text{force}/(\text{length})^2
\stackrel{\rm SI}{=}\text{Pa}$, 
$[c_1,c_2,c_3]=\text{force}\stackrel{\rm SI}{=}\text{N}$ and 
$[d_1,d_2,d_3]=\text{mass}/\text{length}\stackrel{\rm  SI}{=}\text{kg}/\text{m}$. 

The requirement of non-negativity of the energy (material stability) $E=T+W\ge~0$ 
leads to the conditions of semi-positiveness of the constitutive moduli.
Particularly, the constitutive moduli have to fulfill the following conditions~\citep{LA08}
\begin{alignat}{3}
\rho&\ge 0,\nonumber \\
\mu&\ge0,\qquad &\gamma&\ge 0,\qquad &3\lambda+2\mu&\ge 0,\nonumber \\
d_2&\ge 0,\qquad &d_3 &\ge 0,\qquad  &3 d_1+2d_2&\ge 0,\nonumber \\
\label{IE-c}
c_1-c_2&\ge 0,\qquad &c_1+2c_2&\ge 0, \qquad &c_1-c_2+2 c_3&\ge 0.
\end{alignat}

If we substitute the constitutive equations~(\ref{con-eq-3}) and (\ref{con-eq-4}) 
in the equations~(\ref{inhom-di-1}) and (\ref{inhom-di-2}) and use the definitions~(\ref{disl-den}),
we find
\begin{align}
\label{dyn-sys2}
& d_1 ({\dot{\beta}}_{jj,i}-v_{j,ji})
+(d_2+d_3)({\dot{\beta}}_{ij,j}-v_{i,jj})
+(d_2-d_3)({\dot{\beta}}_{ji,j}-v_{j,ji})
+p_i=p^0_i,\\
\label{dyn-sys3}
&d_1\delta_{ij}({\ddot{\beta}}_{kk}-{\dot{v}}_{k,k}) 
+(d_2+d_3)(
{\ddot{\beta}}_{ij}- {\dot{v}}_{i,j})
+(d_2-d_3)({\ddot{\beta}}_{ji}-{\dot{v}}_{j,i})\nonumber\\
&\quad 
+ c_1(\beta_{ik,jk}-\beta_{ij,kk}) 
+c_2(\beta_{ji,kk}-\beta_{jk,ik}+\beta_{kj,ik}-\beta_{ki,jk}) 
+c_3\big[\delta_{ij}(\beta_{lk,lk}-\beta_{ll,kk})\nonumber \\ 
&\quad 
+ (\beta_{kk,ji}-\beta_{kj,ki})\big] 
+\sigma_{ij}
=\sigma^0_{ij},
\end{align} 
which are a coupled system of partial differential equations for the fields
$\Bv$ and $\Bbeta$.

\section{Equations of motion of a screw dislocation} 

We now derive the equations of motion
for an arbitrarily moving screw dislocation. We consider
an infinitely long screw dislocation parallel to the $z$-axis and traveling in
the $xy$-plane.
The symmetry of such a straight screw dislocation leaves only the following non-vanishing 
components of the physical velocity vector and elastic distortion tensor (see,
e.g., \citep{Lardner,Gunther73}):
$v_z$, $\beta_{zx}$, $\beta_{zy}$, and for the dislocation density and
dislocation current tensors: $T_{zxy}$, $I_{zx}$, $I_{zy}$.
The equations of motion of a moving screw dislocation read 
\begin{align}
\label{v-z1}
(d_2+d_3)(\dot{\beta}_{zx,x}+\dot{\beta}_{zy,y}-\Delta v_z)+\rho v_z&=\rho
v^0_z,\\
\label{B-zx1}
(d_2+d_3)(\ddot{\beta}_{zx}-\dot{v}_{z,x})+c_1(\beta_{zy,xy}-\beta_{zx,yy})+(\mu+\gamma)\beta_{zx}&=(\mu+\gamma)\beta^0_{zx},\\
(d_2-d_3)(\ddot{\beta}_{zx}-\dot{v}_{z,x})+c_2(\beta_{zx,yy}-\beta_{zy,xy})+(\mu-\gamma)\beta_{zx}&=(\mu-\gamma)\beta^0_{zx},\\
(d_2+d_3)(\ddot{\beta}_{zy}-\dot{v}_{z,y})+c_1(\beta_{zx,xy}-\beta_{zy,xx})+(\mu+\gamma)\beta_{zy}&=(\mu+\gamma)\beta^0_{zy},\\
\label{B-zy2}
(d_2-d_3)(\ddot{\beta}_{zy}-\dot{v}_{z,y})+c_2(\beta_{zy,xx}-\beta_{zx,xy})+(\mu-\gamma)\beta_{zy}&=(\mu-\gamma)\beta^0_{zy},
\end{align}
where $\Delta=\pd_{xx}+\pd_{yy}$.
In addition, the condition~(\ref{inhom-di}) reads now
\begin{align}
\label{EC}
(\mu+\gamma)(\beta_{zx,x}+\beta_{zy,y})=\rho \dot{v}_z.
\end{align}
From the equations~(\ref{B-zx1})--(\ref{B-zy2}) we obtain the following relations
\begin{align}
\label{Rel-c}
\frac{c_1}{\mu+\gamma}&=-\frac{c_2}{\mu-\gamma} ,\\
\label{Rel-d}
\frac{d_2+d_3}{\mu+\gamma}&=\frac{d_2-d_3}{\mu-\gamma} 
\end{align}
and we may introduce the quantities
\begin{align}
\ell^2_1&=\frac{c_1}{\mu+\gamma},\\
L^2_1&=\frac{d_2+d_3}{\rho},\\
\tau^2_1&=\frac{d_2+d_3}{\mu+\gamma}.
\end{align}
Here $\ell_1$ and $L_1$ are the `static' and `dynamic'  characteristic length
scales 
and $\tau_1$ is the characteristic time scale of the anti-plane strain problem.
Due to the conditions~(\ref{IE-c}), they are non-negative:
$\ell_1\ge 0$, $L_1\ge 0$, $\tau_1\ge 0$.
The velocity of elastic shear waves is defined in terms of the `dynamic'
length scale $L_1$ and the time scale $\tau_1$:
\begin{align}
\label{cT}
c^2_{\TT}=\frac{L_1^2}{\tau_1^2}=\frac{\mu+\gamma}{\rho}.
\end{align}
In a similar way, we introduce the following transversal gauge-theoretical 
velocity defined in terms of 
$\ell_1$ and $\tau_1$:
\begin{align}
\label{aT}
a^2_{\TT}=\frac{\ell_1^2}{\tau_1^2}=\frac{c_1}{d_2+d_3},
\end{align}
and we find the relation
\begin{align}
\label{Rel-a}
\frac{\ell_1^2}{L_1^2}=\frac{a_{\TT}^2}{c_{\TT}^2}.
\end{align}
Applying Eq.~(\ref{EC}), the equations of 
motion~(\ref{v-z1})--(\ref{B-zy2}) can be given in the form
\begin{align}
\label{EOM-v}
\tau_1^2\ddot{v}_{z}-L_1^2\Delta v_z+v_z&=v^0_z,\\
\label{EOM-Bzx}
\tau_1^2\ddot{\beta}_{zx}-\ell_1^2 \Delta \beta_{zx} 
-\tau_1^2 \Big(1-\frac{\ell_1^2}{L_1^2}\Big) \dot{v}_{z,x}
+\beta_{zx}&=\beta^0_{zx},\\
\label{EOM-Bzy}
\tau_1^2\ddot{\beta}_{zy}-\ell_1^2 \Delta \beta_{zy} 
-\tau_1^2\Big(1-\frac{\ell_1^2}{L_1^2}\Big) \dot{v}_{z,y}
+\beta_{zy}&=\beta^0_{zy}.
\end{align}

If we assume that the following relation is valid (see also~\citep{Lazar09})
\begin{align}
\label{L-l}
L_1=\ell_1 ,
\end{align}
then the field
equations~(\ref{EOM-v})--(\ref{EOM-Bzy}) decouple to 
Klein-Gordon equations.
Equations~(\ref{Rel-a}) and (\ref{L-l}) give
the relation $a_{\TT}=c_{\TT}$, that means
that we have only one characteristic velocity $c_{\TT}$ under this assumption.
Particularly, the elastic fields fulfill the uncoupled Klein-Gordon equations
\begin{align}
\label{KGE-v-2}
&\big [1+\ell_1^2 \square_{\TT} \big] v_z=v^0_z,\\
\label{KGE-Bzx-2}
&\big [1+\ell_1^2 \square_{\TT} \big] \beta_{zx} 
=\beta^0_{zx},\\
\label{KGE-Bzy-2}
&\big [1+\ell_1^2 \square_{\TT} \big] \beta_{zy}
=\beta^0_{zy},
\end{align}
with the following $(1+2)$-dimensional d'Alembert operator (wave operator)
\begin{align}
\square_{\TT}&=\frac{1}{c^2_{\TT}}\, \pd_{tt}-\Delta.
\end{align}
In addition, using (\ref{disl-den})
we may derive from (\ref{KGE-v-2})--(\ref{KGE-Bzy-2}) inhomogeneous 
Klein-Gordon equations for the non-vanishing components 
$T_{zxy}$, $I_{zx}$, $I_{zy}$
\begin{align}
\label{KGE-T}
&\big [1+\ell_1^2 \square_{\TT} \big] T_{zxy}=T^0_{zxy},\\
\label{KGE-Ix}
&\big [1+\ell_1^2 \square_{\TT} \big] I_{zx} 
=I^0_{zx},\\
\label{KGE-Iy}
&\big [1+\ell_1^2 \square_{\TT} \big] I_{zy}
=I^0_{zy}.
\end{align}
In field theory, Klein-Gordon equations describe massive fields 
(see, e.g.,~\cite{Rubakov}).
That means that a dislocation is a massive gauge field.

From the condition~(\ref{L-l}), we find for the inertia term of a screw
dislocation
\begin{align}
d_2+d_3=\frac{c_1}{c^2_{\TT}}=\rho\, \ell^2_1,
\end{align} 
that it is given in terms of the characteristic length scale $\ell_1$. 

\section{Nonuniformly moving screw dislocation }

Now we consider a nonuniformly moving 
screw dislocation at the position ($\xi(t),\eta(t)$) at time $t$. The
dislocation line and the Burgers vector $b=b_z$ are parallel to the $z$-axis. 
The dislocation velocity has components: $V_x=\dot{\xi}(t)$, $V_y=\dot{\eta}(t)$.

At first, we want to find the gauge theoretical solutions of the dislocation
density and the dislocation currents for a 
nonuniformly moving screw dislocation. 
Thus, we have to solve the equations~(\ref{KGE-T})--(\ref{KGE-Iy}),
where the right-hand sides are given by the following sources
\begin{align}
\label{DD-0}
T^0_{zxy}&=b\, \delta(x-\xi(t))\delta(y-\eta(t)), \\
\label{Ix-0}
I^0_{zx}&=V_y \, T^0_{zxy},\\
\label{Iy-0}
I^0_{zy}&=-V_x\, T^0_{zxy}\, .
\end{align}
We consider the situation where the source terms have acted after initial 
quiescence at $t\rightarrow -\infty$. 
The solutions for the dislocation density and the dislocation currents are 
the convolution integrals\footnote{
The convolution is defined by:
$T_{zxy}(\Bx,t)=G^{\rm KG}*T^0_{zxy}=
\int_{-\infty}^t \d t'\int\d \Bx' 
G^{\rm KB}(\Bx,t,\Bx',t')\,  T^0_{zxy}(\Bx',t')$.} 
\begin{align}
\label{C-T}
T_{zxy}&=G^{\rm KG}*T^0_{zxy},\\
\label{C-Ix}
I_{zx}&=G^{\rm KG}*I^0_{zx},\\
\label{C-Iy}
I_{zy}&=G^{\rm KG}*I^0_{zy},
\end{align}
where the Green function of the $(1+2)$-dimensional Klein-Gordon equation is defined by
\begin{align}
\big [1+\ell_1^2 \square_{\TT} \big] G^{\rm KG}=\delta(t)\delta(x)\delta(y)\, .
\end{align}
The Green function of the Klein-Gordon equation in $(1+2)$ dimensions is given by~\citep{Iwan}
\begin{align}
\label{G-KG}
G^{\rm KG}
=\frac{1}{2\pi\ell_1^2}\,
\frac{H\big(t-r/c_\TT\big)}{\sqrt{t^2-r^2/c^2_\TT}}\, 
\cos\bigg( \frac{\sqrt{c_\TT^2 t^2-r^2}}{\ell_1}\bigg),
\qquad r^2=x^2+y^2 \, ,
\end{align}
where $H$ is the Heaviside step function.
Substituting~(\ref{DD-0})--(\ref{Iy-0}) and (\ref{G-KG}) into 
(\ref{C-T})--(\ref{C-Iy}), the spatial integration can be performed to
give the results
\begin{align}
\label{T-s}
T_{zxy}&=\frac{b}{2\pi\ell_1^2}\int_{-\infty}^{t_{{\TT}}}\d t'\,
\frac{1}{S_\TT}\,\cos \bigg(\frac{c_\TT S_\TT}{\ell_1}\bigg),\\
\label{Ix-s}
I_{zx}&=\frac{b}{2\pi\ell_1^2}\int_{-\infty}^{t_{{\TT}}}\d t'\,
\frac{V_y(t')}{S_\TT}\,\cos \bigg(\frac{c_\TT S_\TT}{\ell_1}\bigg),\\
\label{Iy-s}
I_{zy}&=-\frac{b}{2\pi\ell_1^2 }\int_{-\infty}^{t_{{\TT}}}\d t'\,
\frac{V_x(t')}{S_\TT}\,\cos \bigg(\frac{c_\TT S_\TT}{\ell_1}\bigg)
\end{align}
with the notations
\begin{align}
\label{Not-s}
&\bar{x}=x-\xi(t')\, ,\qquad \bar{y}=y-\eta(t')\, , 
\qquad \bar{t}=t-t'\, , \qquad
\bar{R}^2=\bar{x}^2+\bar{y}^2, \nonumber\\
&S^2_{\TT}=\bar{t}^2-\frac{\bar{R}^2}{c^2_{\TT}}\, ,
\qquad t_{\TT}={t}-\frac{\bar{R}}{c_{\TT}}\, .
\end{align}
Note that $t_\TT$ is called retarded time, which is the root of the equation
$S_\TT^2=0$ and is less than $t$, and ($\bar{x}, \bar{y})$ is
the distance between the field point $(x,y)$ and the position of
the dislocation $(\xi,\eta)$.
Here, $\xi$ and $\eta$ are the positions in the $x$ and $y$ directions of the
dislocation at time $t'$, $V_x$ and $V_y$ are the velocity components of the
dislocation at the same time $t'$.  
The retarded time $t_\TT$ is the time before $t$, when the dislocation
caused an excitation of the dislocation field, 
which moves from $(\xi,\eta)$ to $(x,y)$ in the time
$\bar{R}/c_\TT$. 
In addition, $t-t_\TT=\bar{R}/c_\TT$ is the time when the dislocation fields move
from $(\xi,\eta)$ to $(x,y)$ with velocity $c_\TT$.
Thus, from each point $\Bx'$, the dislocation fields (\ref{T-s})--(\ref{Iy-s}) draw
contributions emitted at all times $t'$ from $-\infty$ up to $t_\TT$. 
That is the reason why \citet{Eshelby51} said: 
`The dislocation is haunted by its past'. 

\subsection{Solutions for the elastic fields in terms of potential functions}

If we multiply Eqs.~(\ref{KGE-v-2})--(\ref{KGE-Bzy-2}) with $\square_{\TT}$ 
and use the `classical' result~\citep{Eshelby53,Lardner}, 
we obtain
\begin{align}
\label{KGE-v-3}
&\big [1+\ell_1^2 \square_{\TT} \big]\square_{\TT} v_z=I^0_{zx,x}+I^0_{zy,y},\\
\label{KGE-Bzx-3}
&\big [1+\ell_1^2 \square_{\TT} \big]\square_{\TT} \beta_{zx} 
=T^0_{zxy,y}+\frac{1}{c_{\TT}^2} \dot{I}^0_{zx},\\
\label{KGE-Bzy-3}
&\big [1+\ell_1^2 \square_{\TT} \big] \square_{\TT}\beta_{zy}
=T^0_{zyx,x}+\frac{1}{c_{\TT}^2} \dot{I}^0_{zy},
\end{align}
as a set of fourth-order partial differential equations. 
Eqs.~(\ref{KGE-v-3})--(\ref{KGE-Bzy-3}) have the two-dimensional 
form of the Bopp-Podolsky equations~\citep{Bopp,Podolsky} (see also~\citep{Iwan}) 
in generalized electrodynamics, introduced by Bopp and Podolsky in order 
to avoid singularities in electrodynamics.
As source terms only the classical dislocation density $T^0_{zxy}$ and 
dislocation currents $I^0_{zx}$, $I^0_{zy}$ are acting.

Following \citet{Eshelby53}, we express the elastic velocity and the elastic
distortion in terms of `potential' functions ($F$, $A_x$, $A_y$):
\begin{align}
\label{vz-A}
v_z&=A_{y,x}-A_{x,y},\\
\beta_{zx}&=F_{,y}+\frac{1}{c^2_{\TT}}\, \dot{A}_y,\\
\beta_{zy}&=-F_{,x}-\frac{1}{c^2_{\TT}}\, \dot{A}_x.
\end{align}
Then, if $A_x$, $A_y$ and $F$ are chosen to satisfy the subsidiary condition
\begin{align}
A_{x,x}+A_{y,y}+\dot{F}=0,
\end{align}
the field equations~(\ref{KGE-v-3})--(\ref{KGE-Bzy-3}) become equivalent to the 
following inhomogeneous Bopp-Podolsky equations
\begin{align}
\label{KGE-F}
&\big [1+\ell_1^2 \square_{\TT} \big]\square_{\TT} F=T^0_{zxy},\\
\label{KGE-Ax}
&\big [1+\ell_1^2 \square_{\TT} \big]\square_{\TT} A_x=-I^0_{zy},\\
\label{KGE-Ay}
&\big [1+\ell_1^2 \square_{\TT} \big]\square_{\TT} A_y=I^0_{zx}.
\end{align}
The solution of these equations for an infinite medium is defined by
\begin{align}
F&=G^{\rm BP}*T^0_{zxy},\\
A_x&=-G^{\rm BP}*I^0_{zy},\\
A_y&=G^{\rm BP}*I^0_{zx},
\end{align}
where $G^{\rm BP}$ is the Green function of the Bopp-Podolsky equation
(or wave-Klein-Gordon equation)
\begin{align}
\label{BPE}
\big [1+\ell_1^2 \square_{\TT} \big]\square_\TT G^{\rm BP}=\delta(t)\delta(x)\delta(y).
\end{align}
As in generalized 
electrodynamics, we solve~(\ref{BPE}) with the help of two fields
\begin{align}
\label{BPE-2}
\big [1+\ell_1^2 \square_{\TT} \big] G^{\rm BP}=G^\square\, ,\qquad
\square_{\TT}G^\square=\delta(t)\delta(x)\delta(y) 
\end{align}
and we get
\begin{align}
G^{\rm BP}=G^\square-\ell_1^2 G^{\rm KG},
\end{align}
where the first field is the Green function of the two-dimensional wave equation~\citep{Wl,Barton}
\begin{align}
\label{G}
G^\square=\frac{1}{2\pi}\,
\frac{H\big(t-r/c_\TT\big)}{\sqrt{t^2-r^2/c^2_\TT}}
\end{align}
and the second one is the Green function of the Klein-Gordon equation~(\ref{G-KG}).
Finally, we obtain for the Green function of the $(1+2)$-dimensional Bopp-Podolsky
equation 
\begin{align}
\label{G-BP}
G^{\rm BP}
=\frac{1}{2\pi}\,
\frac{H\big(t-r/c_\TT\big)}{\sqrt{t^2-r^2/c^2_\TT}}\, 
\bigg[1-
\cos\bigg( \frac{\sqrt{c_\TT^2 t^2-r^2}}{\ell_1}\bigg)\bigg] \, .
\end{align}

Substituting from (\ref{DD-0})--(\ref{Iy-0}) for the case of a single 
dislocation, the spatial integrations can be performed to give the results
for the potential functions of a screw dislocation
\begin{align}
\label{F}
F&=\frac{b}{2\pi}\int_{-\infty}^{t_{{\TT}}}\d t'\,
\frac{1}{S_\TT}\,
\bigg[1-\cos \bigg(\frac{c_\TT S_\TT}{\ell_1}\bigg)\bigg],\\
\label{Ax}
A_x&=\frac{b}{2\pi}\int_{-\infty}^{t_{{\TT}}}\d t'\,
\frac{V_x(t')}{S_\TT}\,
\bigg[1-\cos \bigg(\frac{c_\TT S_\TT}{\ell_1}\bigg)\bigg],\\
\label{Ay}
A_y&=\frac{b}{2\pi}\int_{-\infty}^{t_{{\TT}}}\d t'\,
\frac{V_y(t')}{S_\TT}\,
\bigg[1-\cos \bigg(\frac{c_\TT S_\TT}{\ell_1}\bigg)\bigg].
\end{align}
In analogy to electromagnetic field theory, we can 
call (\ref{F})--(\ref{Ay}) the (two-dimensional) `Li\'enard-Wiechert
potentials' of a screw dislocation in the gauge theory of dislocations.

\subsection{Radiation part of the elastic fields}
On the other hand, using (\ref{Ix-0}) and (\ref{Iy-0}),  we want to solve directly the 
equations~(\ref{KGE-v-3})--(\ref{KGE-Bzy-3}).
Consequently, we have to solve the equations
\begin{align}
\label{KGE-v-4}
&\big [1+\ell_1^2 \square_{\TT} \big]\square_{\TT} v_z=
V_y\, T^0_{zxy,x}-V_x\, T^0_{zxy,y},\\
\label{KGE-Bzx-4}
&\big [1+\ell_1^2 \square_{\TT} \big]\square_{\TT} \beta_{zx} 
=\frac{\dot{V}_y}{c_\TT^2}\, T^0_{zxy} 
+\Big(1-\frac{V^2_y}{c^2_\TT}\Big)\, T^0_{zxy,y}
-\frac{V_x V_y }{c_{\TT}^2}\, T^0_{zxy,x},\\
\label{KGE-Bzy-4}
&\big [1+\ell_1^2 \square_{\TT} \big] \square_{\TT}\beta_{zy}
=-\frac{\dot{V}_x}{c_\TT^2}\, T^0_{zxy} 
-\Big(1-\frac{V^2_x}{c^2_\TT}\Big)\, T^0_{zxy,x}
+\frac{V_x V_y }{c_{\TT}^2}\, T^0_{zxy,y},
\end{align}
where the right-hand side is given by~(\ref{DD-0}).
It can be seen that only the elastic distortions contain acceleration parts. 
The elastic velocity does not depend on dislocation acceleration.
Using the Green function~(\ref{G-BP}),
we obtain the result for the elastic distortions
\begin{align}
\label{Bzx-s}
\beta_{zx}&=\frac{b}{2\pi c^2_\TT}\Bigg[
\int_{-\infty}^{t_{{\TT}}}\d t'\,\frac{\dot{V}_y(t')}{S_{{\TT}}}\,
\bigg[1-\cos \bigg(\frac{c_\TT S_\TT}{\ell_1}\bigg)\bigg]
+\frac{\pd}{\pd y}
\int_{-\infty}^{t_{{\TT}}}\d t'\,\frac{c^2_\TT-V^2_y(t')}{S_{{\TT}}}\,
\bigg[1-\cos \bigg(\frac{c_\TT S_\TT}{\ell_1}\bigg)\bigg]\nonumber\\
&\qquad
-\frac{\pd}{\pd x}
\int_{-\infty}^{t_{{\TT}}}\d t'\,\frac{V_x(t')V_y(t')}{S_{{\TT}}}\,
\bigg[1-\cos \bigg(\frac{c_\TT S_\TT}{\ell_1}\bigg)\bigg]
\Bigg],\\
\label{Bzy-s}
\beta_{zy}&=-\frac{b}{2\pi c^2_\TT}\Bigg[
\int_{-\infty}^{t_{{\TT}}}\d t'
\,\frac{\dot{V}_x(t')}{S_{{\TT}}}\,
\bigg[1-\cos \bigg(\frac{c_\TT S_\TT}{\ell_1}\bigg)\bigg]
+\frac{\pd}{\pd x}
\int_{-\infty}^{t_{{\TT}}}\d t'\,\frac{c^2_\TT-V^2_x(t')}{S_{{\TT}}}\,
\bigg[1-\cos \bigg(\frac{c_\TT S_\TT}{\ell_1}\bigg)\bigg]\nonumber\\
&\qquad
-\frac{\pd}{\pd y}
\int_{-\infty}^{t_{{\TT}}}\d t'\,\frac{V_x(t')V_y(t')}{S_{{\TT}}}\,
\bigg[1-\cos \bigg(\frac{c_\TT S_\TT}{\ell_1}\bigg)\bigg]
\Bigg].
\end{align}
The solution of Eq.~(\ref{KGE-v-4}) 
for the elastic velocity  has the same expression as given 
in Eq.~(\ref{vz-A}) with (\ref{Ax}) and (\ref{Ay}).
It can be seen that the fields given 
by (\ref{Bzx-s}) and (\ref{Bzy-s}) consist of two parts: 

(a) Field depending on the dislocation velocities $V_x$ and $V_y$ alone and 
proportional to spatial derivatives 
-- velocity field

(b) Field depending on the dislocation accelerations 
$\dot{V}_x$ and $\dot{V}_y$ and proportional to $1/S_\TT$ -- 
acceleration field or radiation field.

All our results depend on $\ell_1$ which is a characteristic length scale 
of the gauge theory of dislocations.


\section*{Acknowledgement}
The author has been supported by an Emmy-Noether grant of the 
Deutsche Forschungsgemeinschaft (Grant No. La1974/1-3). 

\end{document}